\documentstyle[12pt,psfig]{article}
\catcode`\@=11
\long\def\@makefntext#1{
\protect\noindent \hbox to 3.2pt {\hskip-.9pt
$^{{\ninerm\@thefnmark}}$\hfil}#1\hfill}                

\def\@makefnmark{\hbox to 0pt{$^{\@thefnmark}$\hss}}  

\def\ps@myheadings{\let\@mkboth\@gobbletwo
\def\@oddhead{\hbox{}
\rightmark\hfil\ninerm\thepage}
\def\@oddfoot{}\def\@evenhead{\ninerm\thepage\hfil
\leftmark\hbox{}}\def\@evenfoot{}
\def\sectionmark##1{}\def\subsectionmark##1{}}

\renewcommand{\thefootnote}{\fnsymbol{footnote}}

\newcounter{sectionc}\newcounter{subsectionc}\newcounter{subsubsectionc}
\renewcommand{\section}[1] {\vspace*{0.6cm}\addtocounter{sectionc}{1}
\setcounter{subsectionc}{0}\setcounter{subsubsectionc}{0}\noindent
        {\normalsize\bf\thesectionc. #1}\par\vspace*{0.4cm}}
\renewcommand{\subsection}[1] {\vspace*{0.6cm}\addtocounter{subsectionc}{1}
        \setcounter{subsubsectionc}{0}\noindent
        {\normalsize\it\thesectionc.\thesubsectionc. #1}\par\vspace*{0.4cm}}
\renewcommand{\subsubsection}[1]
{\vspace*{0.6cm}\addtocounter{subsubsectionc}{1}
        \noindent
{\normalsize\rm\thesectionc.\thesubsectionc.\thesubsubsectionc.
        #1}\par\vspace*{0.4cm}}

\newcounter{appendixc}
\newcounter{subappendixc}[appendixc]
\newcounter{subsubappendixc}[subappendixc]

\renewcommand{\appendix}[1] {\vspace*{0.6cm}
        \refstepcounter{appendixc}
        \setcounter{figure}{0}
        \setcounter{table}{0}
        \setcounter{equation}{0}
        \renewcommand{\thefigure}{\Alph{appendixc}.\arabic{figure}}
        \renewcommand{\thetable}{\Alph{appendixc}.\arabic{table}}
        \renewcommand{\theappendixc}{\Alph{appendixc}}
        \renewcommand{\theequation}{\Alph{appendixc}.\arabic{equation}}
        \noindent{\bf Appendix \theappendixc #1}\par\vspace*{0.4cm}}

\def\abstracts#1{{

\centering{\begin{minipage}{12.2truecm}\footnotesize\baselineskip=12pt\noindent
        \centerline{\footnotesize ABSTRACT}\vspace*{0.3cm}
        \parindent=0pt #1
        \end{minipage}}\par}}

\renewenvironment{thebibliography}[1]
        {\begin{list}{\arabic{enumi}.}
        {\usecounter{enumi}\setlength{\parsep}{0pt}
\setlength{\leftmargin 1.25cm}{\rightmargin 0pt}
         \setlength{\itemsep}{0pt} \settowidth
        {\labelwidth}{#1.}\sloppy}}{\end{list}}

\newcounter{itemlistc}
\newcounter{romanlistc}
\newcounter{alphlistc}
\newcounter{arabiclistc}

\newcommand{\fcaption}[1]{
        \refstepcounter{figure}
        \setbox\@tempboxa = \hbox{\footnotesize Fig.~\thefigure. #1}
        \ifdim \wd\@tempboxa > 6in
           {\begin{center}
        \parbox{6in}{\footnotesize\baselineskip=12pt Fig.~\thefigure. #1}
            \end{center}}
        \else
             {\begin{center}
             {\footnotesize Fig.~\thefigure. #1}
              \end{center}}
        \fi}

\newcommand{\tcaption}[1]{
        \refstepcounter{table}
        \setbox\@tempboxa = \hbox{\footnotesize Table~\thetable. #1}
        \ifdim \wd\@tempboxa > 6in
           {\begin{center}
        \parbox{6in}{\footnotesize\baselineskip=12pt Table~\thetable. #1}
            \end{center}}
        \else
             {\begin{center}
             {\footnotesize Table~\thetable. #1}
              \end{center}}
        \fi}

\def\@citex[#1]#2{\if@filesw\immediate\write\@auxout
        {\string\citation{#2}}\fi
\def\@citea{}\@cite{\@for\@citeb:=#2\do
        {\@citea\def\@citea{,}\@ifundefined
        {b@\@citeb}{{\bf ?}\@warning
        {Citation `\@citeb' on page \thepage \space undefined}}
        {\csname b@\@citeb\endcsname}}}{#1}}

\newif\if@cghi
\def\cite{\@cghitrue\@ifnextchar [{\@tempswatrue
        \@citex}{\@tempswafalse\@citex[]}}
\def\citelow{\@cghifalse\@ifnextchar [{\@tempswatrue
        \@citex}{\@tempswafalse\@citex[]}}
\def\@cite#1#2{{$\null^{#1}$\if@tempswa\typeout
        {IJCGA warning: optional citation argument
        ignored: `#2'} \fi}}

 1
 1
 1

\font\ninerm=cmr9

\def\Journal#1&#2&#3(#4){#1{\bf #2} (#4) #3.}

\def\NIMA{{\em Nucl. Inst. and Meth. }{\bf A}}

\def\PRD{{\em Phys. Rev. }{\bf D}}
\def\PRC{{\em Phys. Rep. }{\bf C}}

\def\IJMF{{\em Int. J. Mod. Phys. }{\bf A}}

\def\etal{{\it et al.}}

\begin{document}

\centerline{\normalsize\bf MEASUREMENTS OF THE TWO-PHOTON WIDTHS}
\centerline{\normalsize\bf OF THE CHARMONIUM STATES
${\eta_c}$, $\chi_{c{\large{0}}}$ AND $\chi_{c{\large{2}}}$.}
\baselineskip=16pt

\vspace*{0.3cm}
\centerline{\footnotesize VLADIMIR SAVINOV\footnote[1]{
Contribution to the PHOTON95 conference, Sheffield (1995)
\newline
e-mail: savinov@lns62.lns.cornell.edu}
}
\baselineskip=13pt
\centerline{\footnotesize\it School of Physics and Astronomy,
University of Minnesota}
\baselineskip=13pt
\centerline{\footnotesize\it Minneapolis, MN 55455, USA}
\vspace*{0.3cm}
\centerline{\footnotesize ROGER FULTON\footnote[2]{e-mail:
raf@lns62.lns.cornell.edu}
}
\baselineskip=13pt
\centerline{\footnotesize\it Department of Physics,
The Ohio State University}
\baselineskip=13pt
\centerline{\footnotesize\it Columbus, OH 43210, USA}
\vspace*{0.3cm}
\centerline{\footnotesize (representing the CLEO collaboration)}

\vspace*{0.8cm}
\abstracts{Using the CLEO-II detector at the CESR $e^+e^-$ storage ring
running at the center-of-mass energy around 10.6 GeV, we have studied
exclusive production of charmonium states in two-photon collisions.
Employing a dataset comprising 3.0 ${\rm fb}^{-1}$ we have searched for
decays of the $\eta_c$, $\chi_{c0}$ and $\chi_{c2}$ mesons
resulting in four charged hadrons.
We report on the measurements of the cross sections and two-photon widths
of these charmonium states.
}

\normalsize\baselineskip=15pt
\setcounter{footnote}{0}
\renewcommand{\thefootnote}{\alph{footnote}}

\section{Introduction}

In this paper we present the results of a study\cite{ROGER} of
charmonium states exclusively produced in two-photon collisions.
The processes considered here are $e^+e^- \rightarrow e^+e^-R_{c\bar{c}}$,
where $R_{c\bar{c}}$ is
either $\eta_c$, $\chi_{c0}$ or $\chi_{c2}$ mesons.
We measure their production rates when neither the scattered electron
nor the positron is detected (``untagged'' mode).
In these processes a meson is coupled to two space-like
photons one emitted by the electron, the other by the positron.
{}From the measured production rates
we obtain two-photon widths of the mesons.
The following decay channels are used in this study:
$\eta_c \rightarrow K_s^0K^{\pm}\pi^{\mp}$,
$\chi_{c0} \rightarrow \pi^+\pi^-\pi^+\pi^-$ and
$\chi_{c2} \rightarrow \pi^+\pi^-\pi^+\pi^-$.
We also search for the $\eta_c$ decaying into four charged pions.

\section{The Detector}
A brief description of the major components of the CLEO-II
detector\cite{CLEO-II:detector} and its trigger system\cite{CLEO-II:trigger}
can be found in these proceedings.\cite{ff} Additional features of the detector
relevant to this analysis are:
Charged particle identification provided by measurements of
ionization losses in the drift chamber ($dE/dx$) and
time-of-flight measurements from scintillation counters;
Electron identification making use of the $dE/dx$ and $E/p$ ratio, where
$E$ is the energy of the cluster produced by the electron candidate
in the calorimeter and $p$ is the magnitude of the charged track momentum
measured using the drift chambers;
Muon identification uses the depth of particle penetration into the
hadron filter.
This analysis utilizes data recorded with triggers which are designed
to be efficient for events containing multiple charged tracks.

\section{The Event Selection}

In order to select event candidates
we require that the following selection criteria be satisfied.
\begin{itemize}
\item There are exactly four charged tracks
with zero net charge.
\item The root-mean-squared residual of the hits associated with each
track is less than 0.4 mm.
\item The total detected energy is less than 6.0 GeV.
\item The total amount of energy collected in extra cluster(s)\cite{ff}
not associated with any of the charge tracks is less than
500 MeV for the $K_s^0K^\pm\pi^\mp$ mode and less than 650 MeV for the
four charged pions mode.
\item There are no particles identified as electrons, positrons or muons.
\end{itemize}
\noindent
Additional criteria specific to the considered decay channel are discussed
in the next two sections.

\section{The {\large $\eta_c \rightarrow K_s^0K^\pm\pi^\mp$} Decay Channel}

\begin{figure}[h]
\centerline{\psfig{figure=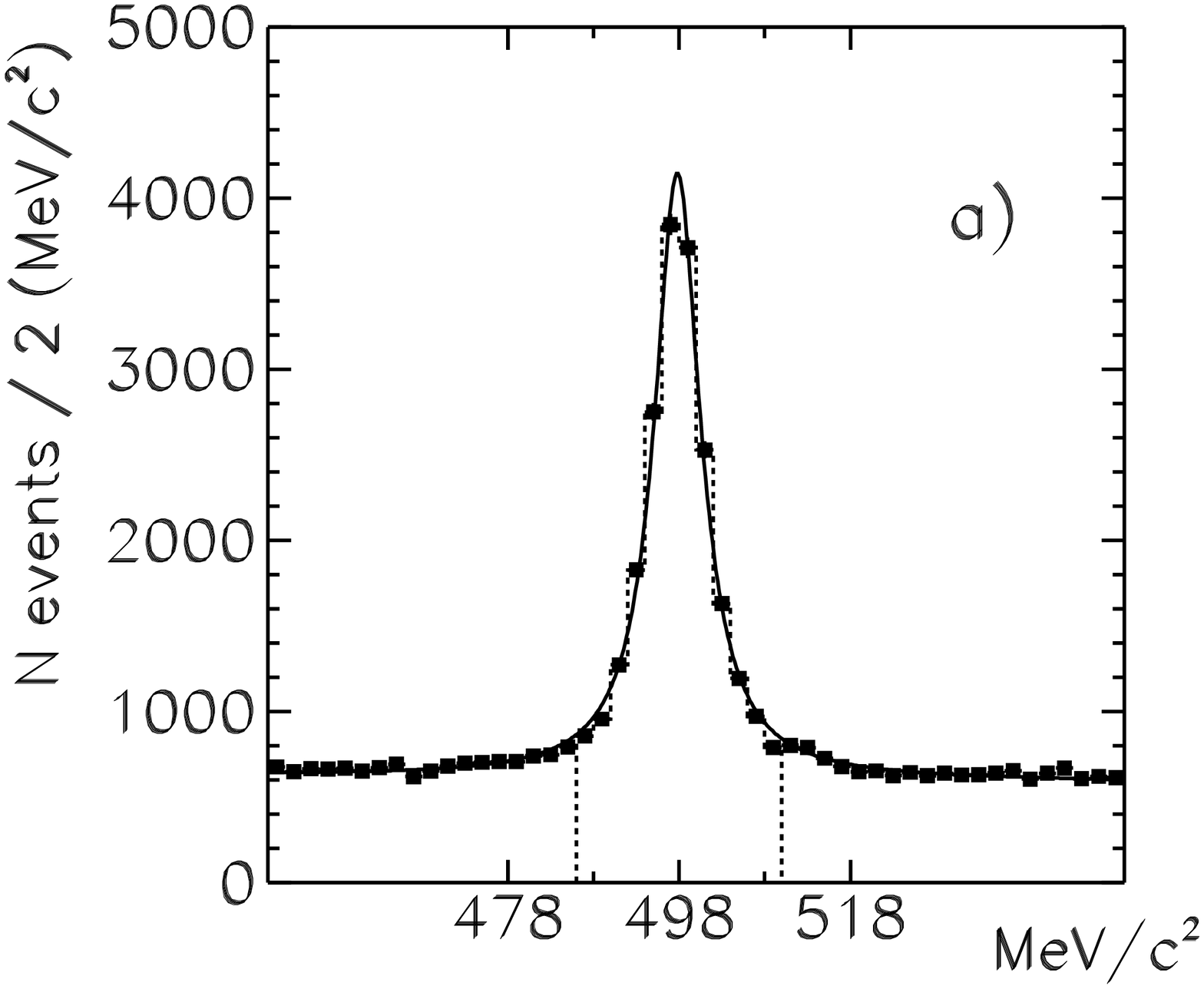,height=1.85in}
	    \psfig{figure=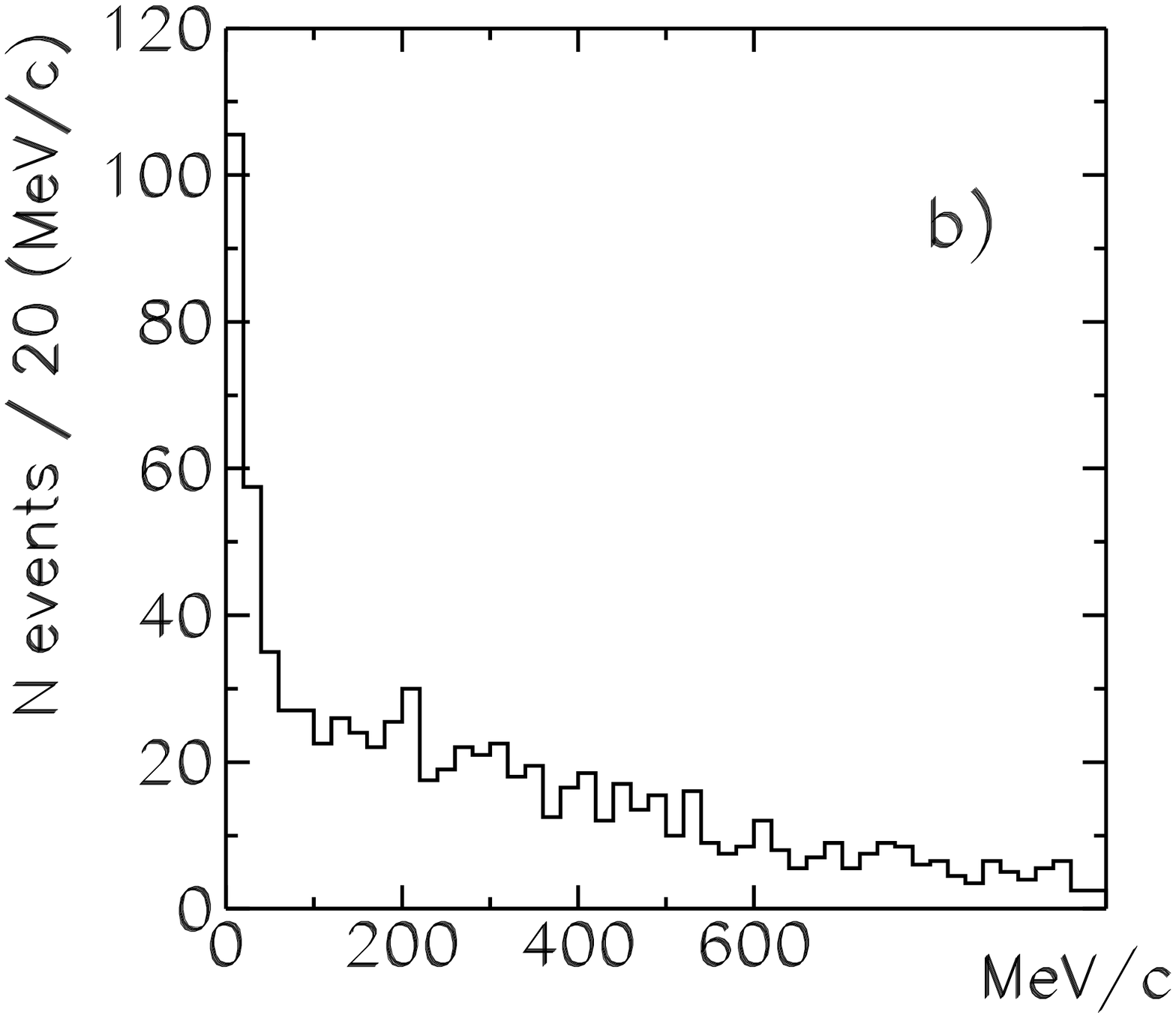,height=1.85in}}
\fcaption{a) The invariant mass spectrum for the $K_s^0$ candidates in
four-prong events.
The curve shows the result of the fit described in the text.
b) The transverse momentum ($p_\perp$) distribution for the
$K_s^0K^\pm\pi^\mp$
candidates with the invariant mass from 2.7 $GeV/c^2$ to 3.3 $GeV/c^2$.}
\label{fig:pt}
\end{figure}

The decay $K_s^0 \rightarrow \pi^+\pi^-$ is identified taking advantage of
the large $K_s^0$ decay length ($c\tau = 2.675$ cm).
We look for pairs of oppositely charged tracks intersecting at a
transverse distance ({\it i.e.} projected distance
in the plane perpendicular to the beam axis)
greater than 1 mm from the primary interaction point.
The momentum of each pair is calculated using
the track momenta evaluated at the secondary vertex.
The energy is calculated assuming that
tracks are produced by pions.
The invariant mass of each pair is then calculated.
The resulting mass spectrum in Fig.\ref{fig:pt}a shows clear evidence of
$K^0_s$ decays\footnote[1]
{The position of the peak in the mass distribution is obtained by fitting
this distribution with a Breit-Wigner function for the signal and
a constant for the background.}.
{}~A pair is accepted as a $K_s^0$ candidate
if its invariant mass falls within 12.0 MeV/${\rm c}^2$ of the peak position.
We reject events if the two remaining tracks are consistent
with a $K_s^0$ decay:
{\it i.e.} if their invariant mass falls within 12.0 MeV/${\rm c}^2$
of the peak position of the previously discussed distribution.
To supress beam-gas background events and to select events of
reasonable quality we require that the two tracks recoiling against
the $K_s^0$ come from the
nominal interaction region: their closest approach
to the primary interaction point is less than 50 mm in the longitudinal
direction and less than 5 mm in the transverse plane.
These tracks are assigned the $K$ and $\pi$ masses
and the $K_s^0K^\pm\pi^\mp$ mass value is calculated for each of the
two possible mass assignments.
The upper tail probability of the {$\chi^2$} distribution\cite{UTP}
($P_{{\rm utp}}$)
is calculated for each mass assignment using TOF and $dE/dx$ information.
In the ideal case, the correct choice of particle combinations
produces a uniform $P_{{\rm utp}}$ distribution
while incorrect particle combinations tend to congregate near zero.
To supress unwanted background events which do not contain charged kaons
$P_{{\rm utp}}$ is required to be greater than 0.05.
Most candidates fail this condition while
the remaining events are dominated by a single acceptable mass assignment.
Events are given a net weight of one.
For events with two acceptable combinations, each possibility is given a
weight proportional to its value of $P_{{\rm utp}}$.

In the two-photon untagged events substantial energy and momentum
are carried away by the electron and positron scattered at small angles.
However, since the transverse momenta of scattered electrons are small,
that of the hadronic system should also be small.
Hence, the distribution of the transverse momentum
of the hadronic system ($p_\perp$) peaks at zero as shown in Fig.\ref{fig:pt}b.
Events with a transverse momentum ($p_\perp$) of the hadronic system
less than 200 MeV/c
are accepted into the final invariant mass plot for the
$K_s^0K^\pm\pi^\mp$ candidates which is shown in Fig.\ref{fig:signals}a.
To obtain the number of events in the signal peak we fit this distribution
with a signal shape function\footnote[1]
{The signal shape function is a combination of two Gaussians with different
widths and the same peak position.
It reflects the variation of the detector resolution over the angular
and energy ranges of the decay products.}
{}~obtained by running a GEANT-based detector simulation program.
A smooth background contribution is approximated by a power function
of the form $N(W)~=~N_0 W^x$, where $N_0$ is a normalization parameter,
$x$ is a power parameter and $W$ is the invariant mass.
The number of observed events is shown in Table \ref{tab:table_signal}.

\section{The $\pi^+\pi^-\pi^+\pi^-$ Decay Channel}

To identify the decay mode into four charged pions we require that all
charged tracks originate from the interaction region.
\begin{figure}[h]
\centerline{\psfig{figure=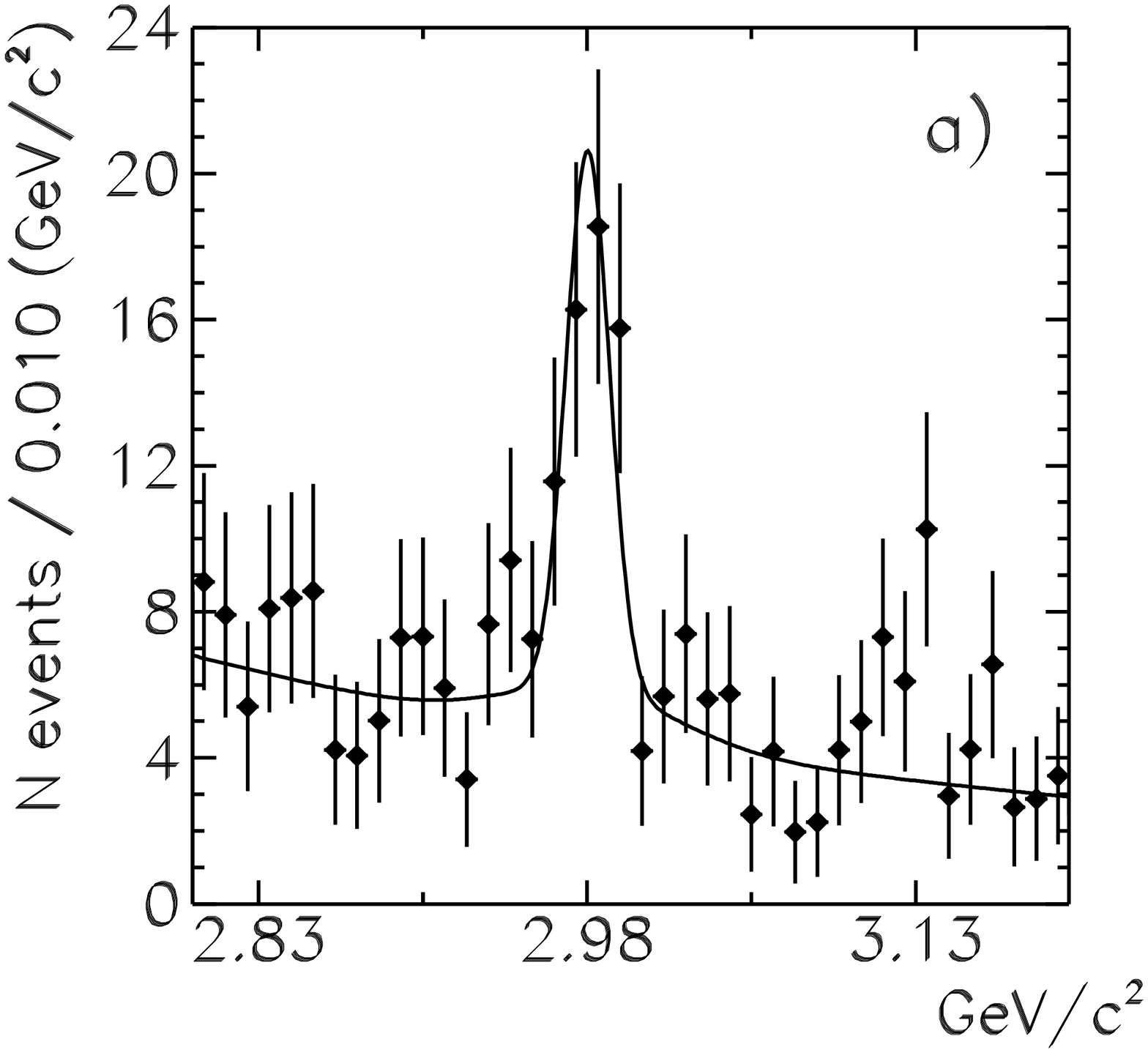,height=1.75in}
	    \psfig{figure=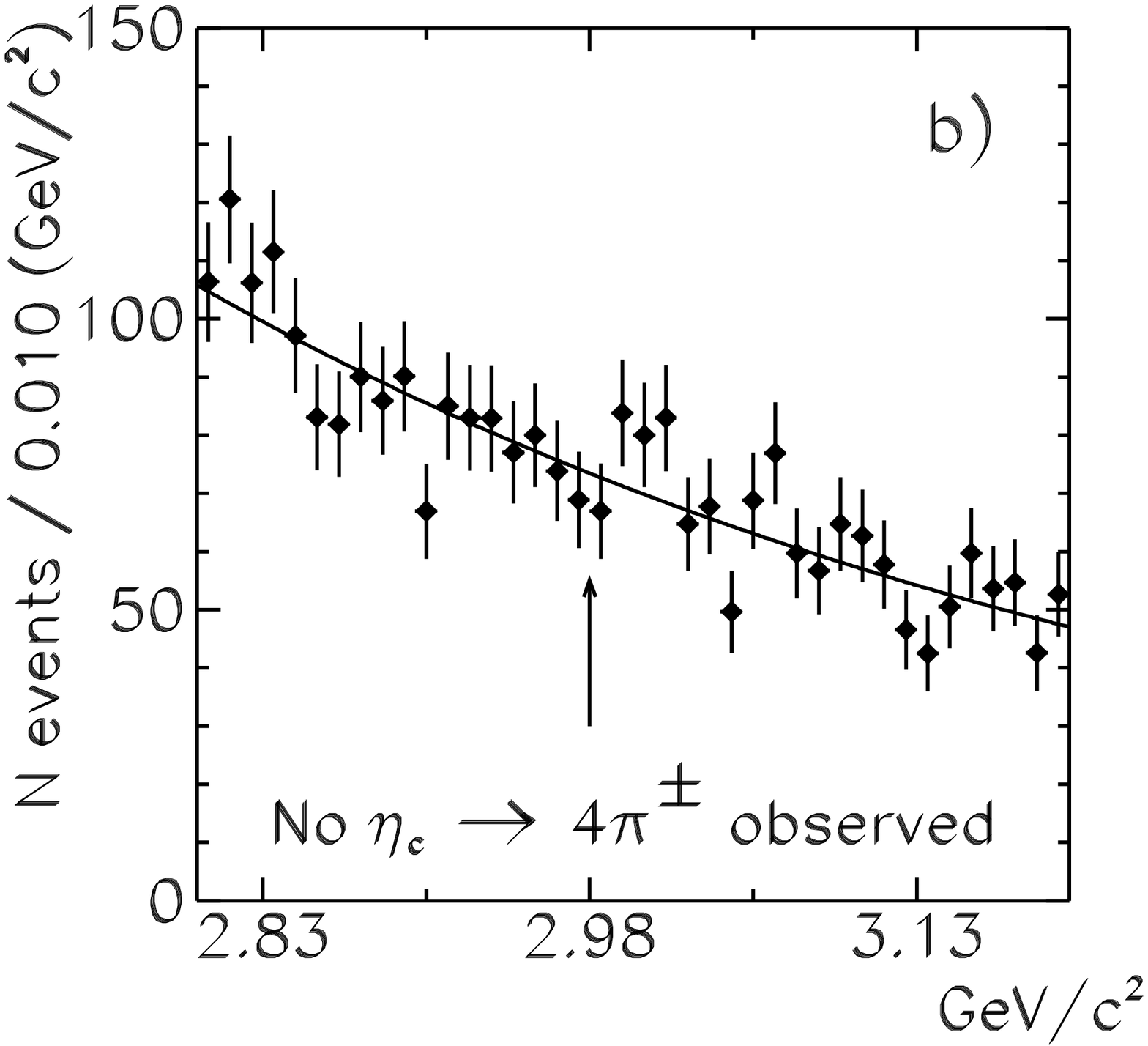,height=1.75in}
	    \psfig{figure=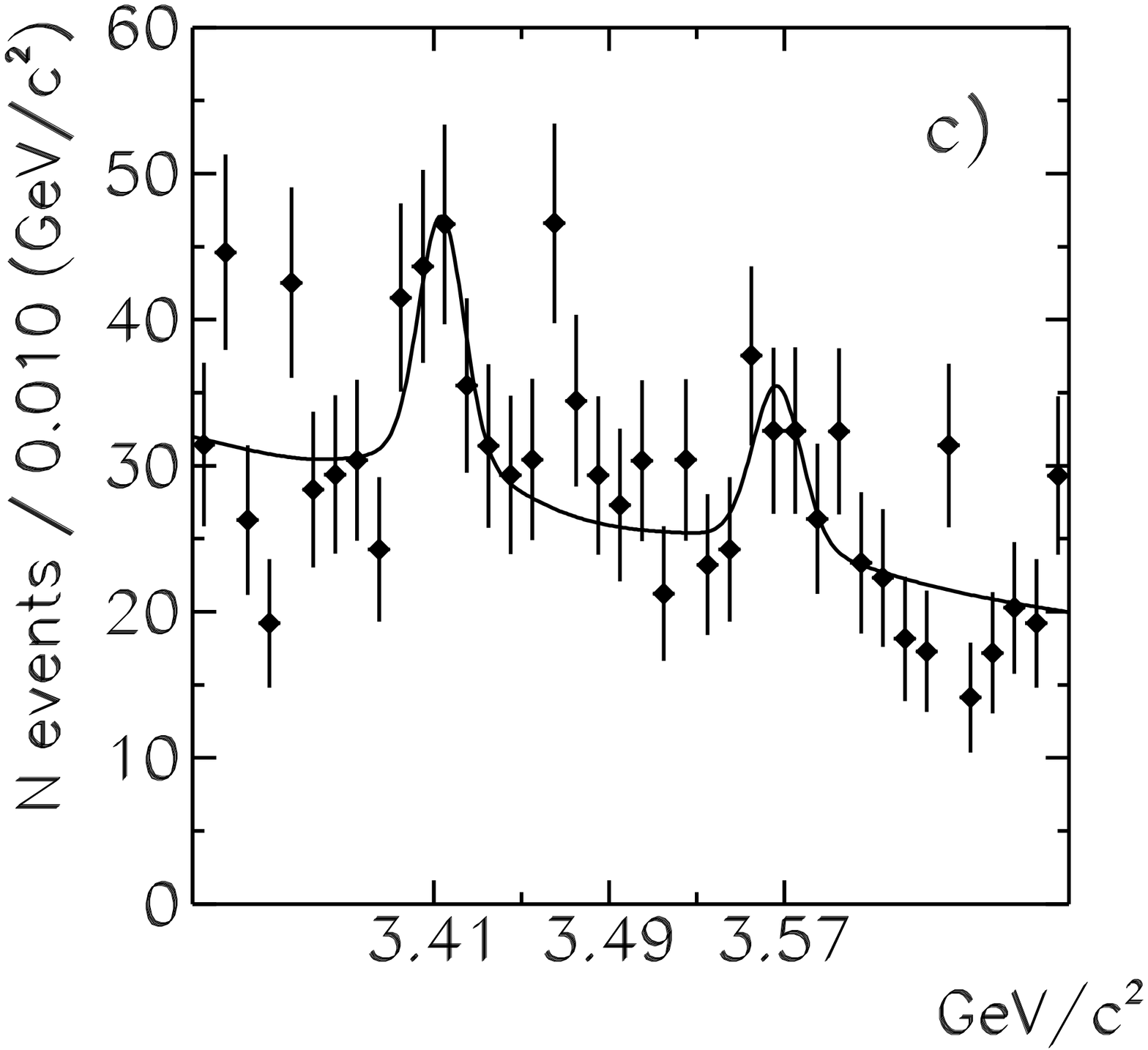,height=1.75in}}
\fcaption{The mass distributions for the candidates: a) $K_s^0K^\pm\pi^\mp$,
b) $\pi^+\pi^-\pi^+\pi^-$ for $\eta_c$ mass range and
c) $\pi^+\pi^-\pi^+\pi^-$ for $\chi_c$ mass range.
The curves show the results of the fits described in the text.}
\label{fig:signals}
\end{figure}
To satisfy this requirement each track must have a separation from the
beam crossing in the longitudinal direction of less than 50 mm and in
the transverse plane of less than 5 mm.
The upper tail probability of the {$\chi^2$} distribution is determined
from TOF and $dE/dx$ measurements assuming all four tracks to be pions.
Candidates with $P_{{\rm utp}}$ greater than 0.1 are accepted into
the $\pi^+\pi^-\pi^+\pi^-$ invariant mass plot.
In addition, we require that $p_\perp$ of the hadronic system
be less than 200 MeV/c.  Signals for both the $\chi_{c0}$ and
$\chi_{c2}$ are seen in Fig.\ref{fig:signals}c.
The numbers of signal events are obtained in the same way as
in the previous section.
The background contribution is approximated by an exponential function.

We do not observe $\eta_c$ in the decay mode into four charged pions.
The distribution of the invariant mass around the region where
one expects $\eta_c$ is shown in Fig.\ref{fig:signals}b.
The vertical arrow shows where an $\eta_c$ signal is expected.

\section{Results}

We obtain the detection efficiency by running a
detector simulation program which includes trigger simulation.
The trigger is efficient for 70\% of all events and for 94\% of events
which would satisfy analyses selection criteria.
These efficiencies agree within 5\% with trigger efficiencies measured
in data using redundancies between different triggers.
\begin{table}
\begin{center}
\tcaption{Summary of $\eta_c$, $\chi_{c0}$ and $\chi_{c2}$ measurements.}
\smallskip
\smallskip
\label{tab:table_signal}
\begin{tabular}{lccccc}
\hline
\hline
        & Number  	& Branching     &
Br $\cdot~\Gamma_{\gamma\gamma}$	& $\Gamma_{\gamma\gamma}$	\\
Channel & observed	&
Fractions(\%)\cite{PDG:94} & (0.01
$\times$ KeV)			& (KeV)			 	\\
\hline
$\eta_c \rightarrow$	& $54.1 \pm 12.6$	&
$1.5 \pm 0.4$ & $6.5 \pm 1.5 \pm 1.1$ & $4.3 \pm 1.0 \pm 0.7 \pm 1.4$ \\
$K_s^0K^\pm\pi^\mp$	& 	& & & \\
$\eta_c \rightarrow$	&               	&
$1.2 \pm 0.4$ & $< 2.7 ~( ~90\% ~{\rm CL}~)$ & $< 2.4 ~( ~90\% ~{\rm CL~})$ \\
$\pi^+\pi^-\pi^+\pi^-$	& 	& & & \\
$\chi_{c0} \rightarrow$	& $47.2 \pm 15.5$	& $3.7 \pm 0.7$ &
$6.4 \pm 2.1 \pm 1.5$ & $1.7 \pm 0.6 \pm 0.4 \pm 0.3$ \\
$\pi^+\pi^-\pi^+\pi^-$	&	& & & \\
$\chi_{c2} \rightarrow$	& $41.9 \pm 13.8$	& $2.2 \pm 0.5$ &
$1.5 \pm 0.5 \pm 0.3$ & $0.7 \pm 0.2 \pm 0.1 \pm 0.2$ \\
$\pi^+\pi^-\pi^+\pi^-$	&	& & & \\
\hline
\end{tabular}
\end{center}
\end{table}
\noindent
For simulation of the two-photon processes we use
Monte Carlo generators based on the formalism of Budnev.\cite{Budnev}
{}~The intrinsic properties of the charmonium resonances are described by
a double pole form factor model.\cite{Poppe}
{}~The pole mass has been varied from zero to the $J/\psi$ mass.
We estimate that the uncertainty in the pole mass and form factor model
are sources of a relative systematic error of 10\%.
Other sources of systematic errors include uncertainties in the
detection efficiency due to the
event selection criteria, trigger and detector simulation and tracking.
Overall detection efficiencies are 12\%, 20\% and 21\% for the
$\eta_c \rightarrow K_s^0K^\pm\pi^\mp$,
$\chi_{c0} \rightarrow 4\pi^\pm$ and $\chi_{c2} \rightarrow 4\pi^\pm$  decay
channels, respectively.

The results of this study are summarized in Table \ref{tab:table_signal}.
The first two errors quoted are statistical and systematic.
The third error in the two-photon widths arises from
the uncertainties in the corresponding branching fractions.
Our measurement of the $\Gamma_{\gamma\gamma}(\eta_c)$ = 4.3 KeV
using the $\eta_c \rightarrow K_s^0K^{\pm}\pi^{\mp}$ decay mode
is consistent with previous measurements which are
in the 5-28 KeV range,
though most recent and more precise measurements fall below
10 KeV.
The upper limit on $\Gamma_{\gamma\gamma}(\eta_c)$ derived from the decay
$\eta_{c} \rightarrow \pi^+\pi^-\pi^+\pi^-$ is
$\Gamma_{\gamma\gamma}(\eta_c) < 2.4$ KeV at the 90\% confidence
level.
Our measurements of $\Gamma_{\gamma\gamma}$
for ${\chi_{c0}}$ and $\chi_{c2}$ mesons are consistent
with results of previous experiments.\cite{PDG:94}

\section{Discussion and Summary}

The nonrelativistic quark model predicts that the two-photon width
of the bound $q\bar{q}$ state with angular momentum {\it l}
{}~is proportional to the square of the {\it l}-th derivative
of the ${\it q\bar{q}}$ pair wavefunction at the origin
({\it i.e.} at contact).
To calculate the two-photon width one needs to know this wave function,
or in terms of the Fourier transform, the momentum distribution
of the ${\it q\bar{q}}$ pair inside the meson. This momentum distribution
can be obtained from the nonrelativistic potential model.
However, due to this non-perturbative calculation the overall normalization
of the width is not very reliable.
This uncertainty can be partially removed when one considers the ratio of
two partial widths of the same meson or similar mesons.
The ratio of the widths may be calculated more accurately
using a relativistic approach.

For the $\eta_c$ meson a nonrelativistic model\cite{ROSNER}
with next-to-leading order quantum chromodynamics corrections predicts:
$
\Gamma_{\gamma\gamma}(\eta_c)/\Gamma_{\mu^+\mu^-}(J/\psi) = (4/3)
(1 + 1.96 \alpha_s/\pi).
$
Using the measured width
$\Gamma(J/\psi \rightarrow \mu^+\mu^-) \approx 5.3$ KeV\cite{PDG:94}
with a choice of the $\alpha_s \approx 0.28$,
this yields $\Gamma(\eta_c \rightarrow \gamma\gamma) \approx 8.2$ KeV.
This calculation assumes that the $\eta_c$ and $J/\psi$ wavefunctions are
identical at the origin, ignoring hyperfine mass splitting and
coupled-channel effects. Indeed, the hyperfine effects would be likely
to enhance the wavefunction of the $\eta_c$ relative to that of the
$J/\psi$.\cite{ROSNER} ~Some evidence that the
wavefunctions are not identical comes from the supression of the
$J/\psi \to \gamma \eta_c$ magnetic transition.
On the other hand a
description which incorporates relativistic effects\cite{BARNES_1}
with a wavefunction obtained from a potential model
predicts $\Gamma_{\gamma\gamma}(\eta_c) \approx 4.8$ KeV.\footnote[1]
{This is evaluated assuming the mass of the {\it c} quark ($m_c$) to be
1.4 GeV. A somewhat larger $m_c$ = 1.6 GeV leads to the
$\Gamma_{\gamma\gamma}(\eta_c)$ = 3.4 KeV. The choice of the {\it c} quark mass
becomes even more important for {\it P}-wave $q\bar{q}$ states because of the
higher power of the {\it c} quark mass involved.}
{}~This calculation agrees with our measurement of the
$\Gamma_{\gamma\gamma}(\eta_c)$ = $4.3 \pm 1.0 \pm 0.7 \pm 1.4$ KeV
obtained in the $K_s^0K^{\pm}\pi^{\mp}$ mode.

For the $\chi_{c0}$ and $\chi_{c2}$ states a naive nonrelativistic quark model
predicts $\Gamma_{\gamma\gamma}(\chi_{c0})$ :
$\Gamma_{\gamma\gamma}(\chi_{c2})$ =
15/4 : 1. The next-to-leading order quantum chromodynamics
corrections\cite{ROSNER} change this ratio by a
factor of $(1+0.2 \alpha_s/\pi)/(1-16\alpha_s/(3\pi))$.
By evaluating this expression at $\alpha_s \approx 0.28$ we obtain
$\Gamma_{\gamma\gamma}(\chi_{c0})/\Gamma_{\gamma\gamma}(\chi_{c2})
\approx 7.3$.
The relativistic calculation\cite{BARNES} predicts this number to be
around 2.8.
Only the latter value is consistent with our measurement of $2.4 \pm 1.1
\pm 0.7$.

\end{document}